\documentclass{llncs}
\usepackage{graphicx,float}
\input{psfig.sty}
\begin{document}
\pagestyle{empty}
\mainmatter

\title{Contagion Flow Through Banking Networks }
\titlerunning{Contagion flow in banking networks}
\author{Michael Boss$^{1}$, Martin Summer$^{1}$, Stefan Thurner$^{2}$}
\authorrunning{Boss, Summer, Thurner}
\institute{ $^{1}$ Oesterreichische Nationalbank,  
Otto-Wagner-Platz 3, A-1011 Wien, Austria 
\footnote{
The views and findings of this paper are entirely those 
of the authors and do not necessarily represent the views 
of Oesterreichische Nationalbank.}\\
$^{2}$ Complex Systems Research Group, HNO, Medical University of Vienna, \\
W\"ahringer G\"urtel 18-20,  A--1090 Vienna, Austria\\
\email{thurner@univie.ac.at}  
}
\maketitle
\begin{abstract}
Starting from an empirical analysis of the network structure 
of the Austrian inter-bank market,  
we study the flow of funds through the 
banking network following exogenous shocks to the system. 
These shocks are implemented by stochastic changes in 
variables like interest rates, exchange rates, etc.    
We demonstrate that the system is relatively stable 
in the sence that defaults of individual banks are unlikely 
to spread over the entire network. 
We study the contagion impact of all individual banks, meaning
the number of banks which are driven into insolvency as a result 
of a single bank's default. We show that the vertex betweenness of 
individual banks is linearly related to their contagion impact. 
\end{abstract}

\section{Introduction}

With the development of new ideas and tools of analysis, 
physics has lately strongly contributed to a functional 
understanding of the structure of complex real world networks.  
A key insight of this research has been the discovery of surprising 
structural similarities in seemingly very different networks, ranging from 
internet technology to cell biology.    
See \cite{dormen03} for an overview. 
Remarkably, many real world networks 
show  power-law (scale free) degree distributions and
feature a certain pattern of cliquishness, 
quantified by the  {\it clustering coefficient}.  Many also 
exhibit a so 
called {\it small world phenomenon}, meaning that  
the average shortest path between any two vertices 
(''degrees of separation'') in the network can be surprisingly small.  
Maybe one of the most important contributions to recent network 
theory is an interpretation of these network parameters with respect 
to stability, robustness, and efficiency of an underlying system. 
From this perspective financial networks are a natural 
candidate to study. 
The financial world can be thought of as a set of intermediaries 
i.e. banks who interact with each other through  
financial transactions. These interactions are governed by a set 
of rules and regulations, and take place on an interaction 
graph of all connections between financial intermediaries.  
Financial crises i.e. the incapacity to finance businesses and 
industries have recently hit several countries all around the globe. 
These events have triggered a boom of papers on banking-crises, 
financial risk-analysis and  numerous policy 
initiatives to find and understand the weak-points of the financial system. 
One of the major concerns in these debates is the danger of so 
called {\it systemic risk}: the large scale breakdown of financial 
intermediation due to domino effects of insolvency \cite{hardeb00,sum03}.
The network of mutual credit relations between 
financial institutions is supposed to play a key role
in the risk for contagious defaults. 

In the past the theoretical economic literature
on contagion \cite{algal00,thurner03} suggest
network topologies that might be interesting to look at.
In \cite{algal00}  it is suggested to study a complete
graph of mutual liabilities. The properties of a banking system with
this structure is then  compared to properties of 
systems with non complete networks.
In \cite{thurner03} a 
set of different network structures is studied.
However, so far surprisingly little is known 
about the {\it actual} empirical network topology of mutual
credit relations between financial institutions. 

In a recent paper we have for the first time 
analyzed empirical data to reconstruct the banking 
network of Austria \cite{boss03}. 
Here we  characterized the interbank network by the 
liability (or exposure)  matrix $L$. The entries 
$L_{ij}$ are the liabilities bank $i$ 
has towards bank $j$.  
$L$ is a square matrix but not necessarily symmetric. 
We showed that the liability ($L_{ij}$) size distribution follows 
a power law, which can be understood as being driven by underlying 
size and wealth distributions of the banks, which show similar 
power exponents.  
We find that the interbank network shows -- like many other realistic   
networks  -- power law dependencies in the
degree distributions.  
We could show that different scaling exponents within the 
same distribution relate to different hierarchy levels in 
sub-networks (sectors) within the total 
network. The scaling exponents by the agricultural banks are very low, 
due to the hierarchical structure of this sector, while the other banks 
lead to scaling exponents of sizes 
also found in other complex real world networks. 
The interbank network shows a low
clustering coefficient, a result that mirrors the 
analysis of community structure which shows
a clear network pattern, where banks 
would first have links with their 
head institution, whereas these few head institutions 
hold links among each other. A consequence 
of this structure is that the interbank network is a 
small world with a very low ''degree of separation'' 
between any two nodes in the system. 

\section{Inter-Bank Topology and Flow of Contagion}
The knowledge of the detailed structure of the interbank topology 
enables us to use it as an input for a contagion-flow model 
\cite{elsinger02}, which is the main idea of this paper.
Unlike previous research where we studied random flow on structured 
networks \cite{TT,TTR}, here we follow the flow of payments 
uniquely determined by the 
structure of the liability matrix. 
We use this  flow to  perform stress tests 
to the system by artificially changing external financial 
parameters like interest rates, exchange rates etc., mimicking conditions 
and ''global'' events 
which are beyond the influence of banks. 
By doing this we can  
answer questions about the stability of the financial system 
with respect to external shocks, in particular, 
which banks are 
likely to default due to shocks, and which banks will 
drag other banks into default due to their mutual credit 
relations (contagion).

In the following we are looking for the bilateral 
liability  matrix $L$ of all (about $N=900$) Austrian banks, 
the Central Bank (OeNB) and an aggregated foreign banking 
sector. Our data consists of 10 $L$ matrices each representing 
liabilities for quarterly single month periods between 
$2000$ and $2003$.  T obtain these data,  
we draw upon two major sources from the Austrian Central Bank: 
the Austrian bank balance sheet data base (MAUS) and the 
major loan register (GKE). 
The Austrian  banking system has a sectoral organization 
due to historic reasons. Banks belong to one of seven sectors: 
savings banks (S),
Raiffeisen (agricultural) banks (R), 
Volksbanken (VB), 
joint stock banks (JS), 
state mortgage banks (SM), 
housing construction savings and loan associations (HCL), and 
special purpose banks (SP). 
Banks have to break down their balance sheet
reports on claims and liabilities with other banks according 
to the different banking sectors, Central Bank and 
foreign banks. This practice of reporting on balance interbank
positions breaks the liability matrix $L$ down to blocks of sub-matrices for
the individual sectors. 
Banks with a head institution have to disclose their
positions with the head institution, which gives additional 
information on $L$. Since many banks in the system hold interbank 
liabilities only with their head institutions,  one 
can pin down many entries in the $L$ matrix exactly. 
This information is combined with the data 
from the major loans register of OeNB. This 
register contains all interbank loans above a threshold of 360 000 Euro. 
This information provides us with a set of constraints 
and zero restrictions for individual entries $L_{ij}$.
Up to this point one can obtain about 
$90\%$ of the $L$-matrix entries exactly. 
For the rest we employ entropy maximization method 
\cite{boss03}. 
The estimation problem can be set up as a standard convex 
optimization problem:
Assume we have a total of $K$ constraints. 
The column and row constraints take the form 
$ \sum_{j=1}^{N}L_{ij}=b_i^r \quad \forall \quad i $ 
and $\sum_{i=1}^{N}L_{ij}=b_j^c \quad \forall \quad j $
with $r$ denoting {\it row} and $c$ denoting {\it column}. 
Constraints imposed by the knowledge about particular entries in $L_{ij}$ 
are given by
$b^l \leq L_{ij} \leq b^u$ for some $i,j$
The aim is to find the matrix $L$ 
that has the least discrepancy to some a priori matrix $U$ with respect 
to the (generalized) cross entropy measure 
\begin{equation}
{\cal C}(L,U)=\sum_{i=1}^{N}\sum_{j=1}^{N}L_{ij}\ln 
\left(\frac{L_{ij}}{U_{ij}}\right)  \quad . 
\label{zielf}
\end{equation}
$U$ is the matrix which contains all known exact liability 
entries. For  those entries (bank pairs) ${ij}$ where we have 
no knowledge from Central Bank data, we set $U_{ij}=1$. 
We use the convention that $L_{ij}=0$ whenever $U_{ij}=0$ and 
define $0\ln (\frac{0}{0})$ to be $0$. 
As a result we obtain a rather precise  
picture of the interbank relations at a particular point in time. 
Given $L$ we find that the distribution of liabilities 
follows a power law for more than three decades with an 
exponent of $-1.87$ \cite{boss03}.
To extract the network topology from these data, 
for the present purposes we ignore directions and 
define an undirected but weighted 
adjacency matrix $A^w_{ij}=L_{ij}+L_{ji}$, which measures the gross 
interbank interaction, i.e. the total volume of 
liabilities and assets for each node. 
We next test the validity of our estimate of $L$, 
by computing the implied community structure and then 
comparing it to the known real bank-clusters, i.e. the sectors. 
There exist various ways to find functional clusters 
within a given network. 
In \cite{zhou03_a} an algorithm was introduced 
which -- while having at least the same performance 
rates as \cite{girvan02} -- provides an additional measure for 
the differences of  cluster, 
the so-called dissimilarity index. 
For analyzing our interbank network we apply this latter 
algorithm to the weighted adjacency matrix $A^w$. 
As the only preprocessing step we clip all entries in 
$A^w$ above a level of 300 m Euro for numerical reasons, i.e. 
$A^w_{\rm clip} = \min(A^w,300 {\rm m})$. 
The  community structure obtained in this way  
can be compared to the actual community structure in the real 
world.  
\begin{figure}[t]
\begin{center}
\begin{tabular}{ccc} 
\hspace{-3mm}\includegraphics[height=30mm]{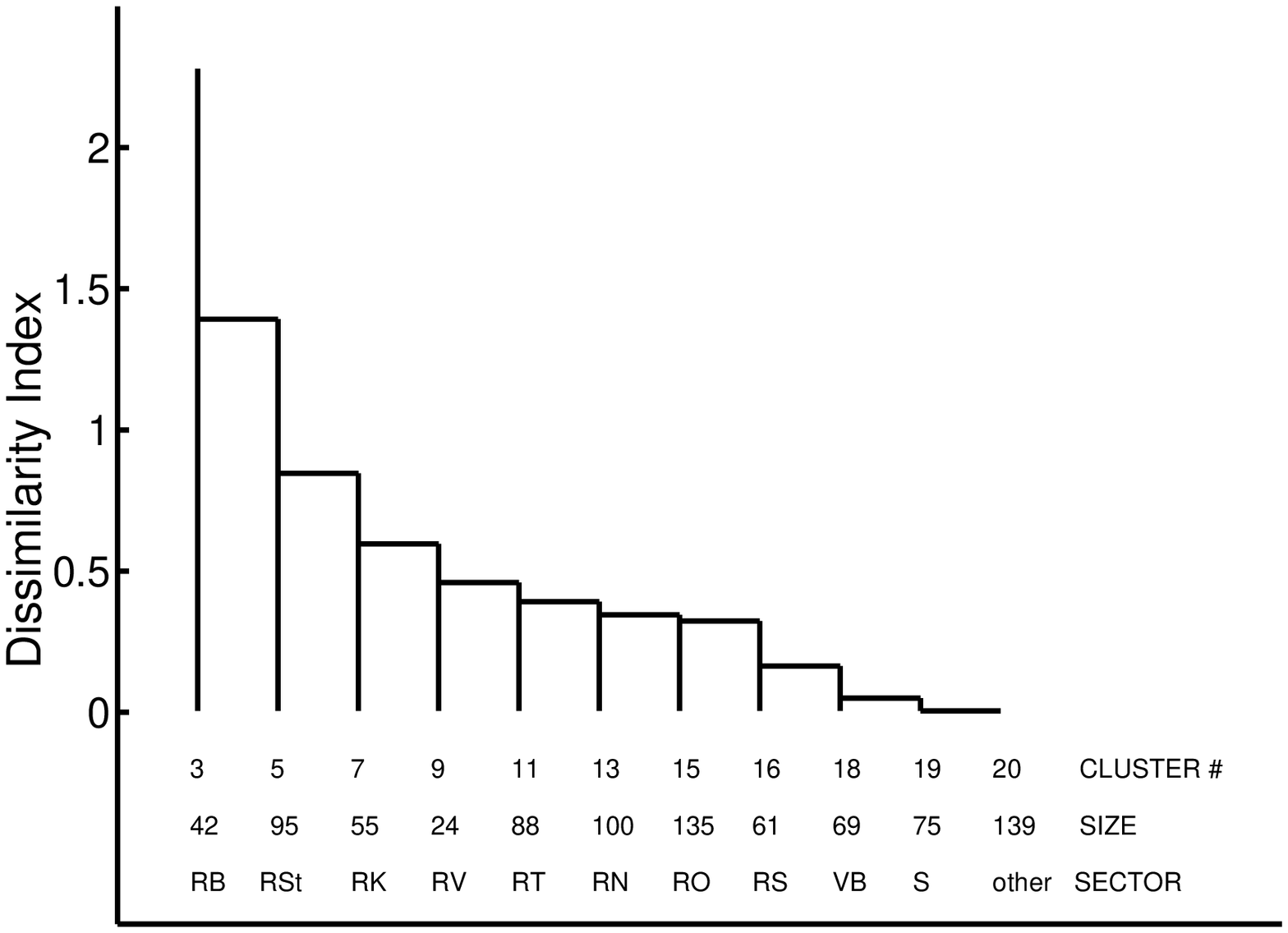} &
\includegraphics[height=32mm]{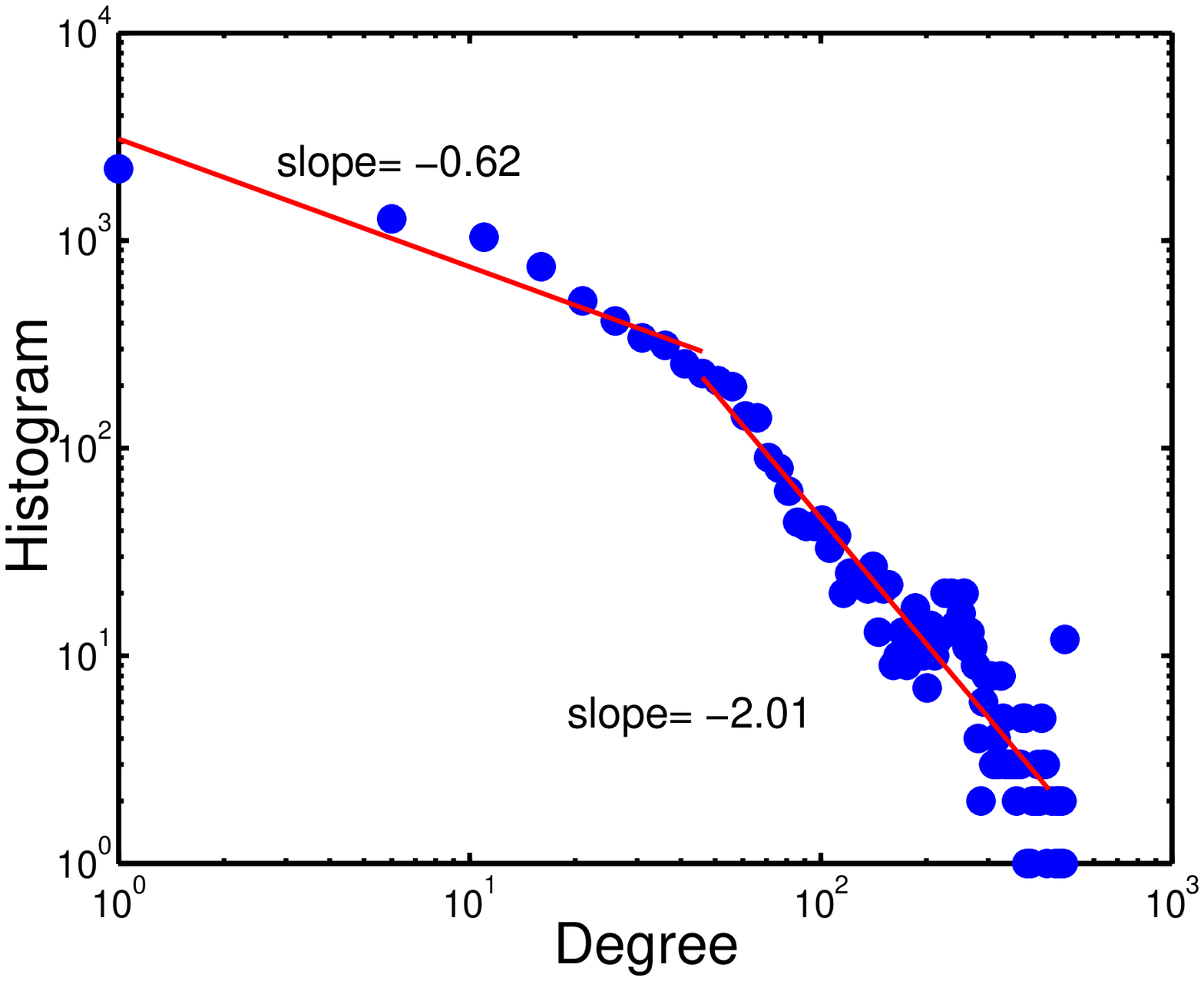} & 
\includegraphics[height=32mm]{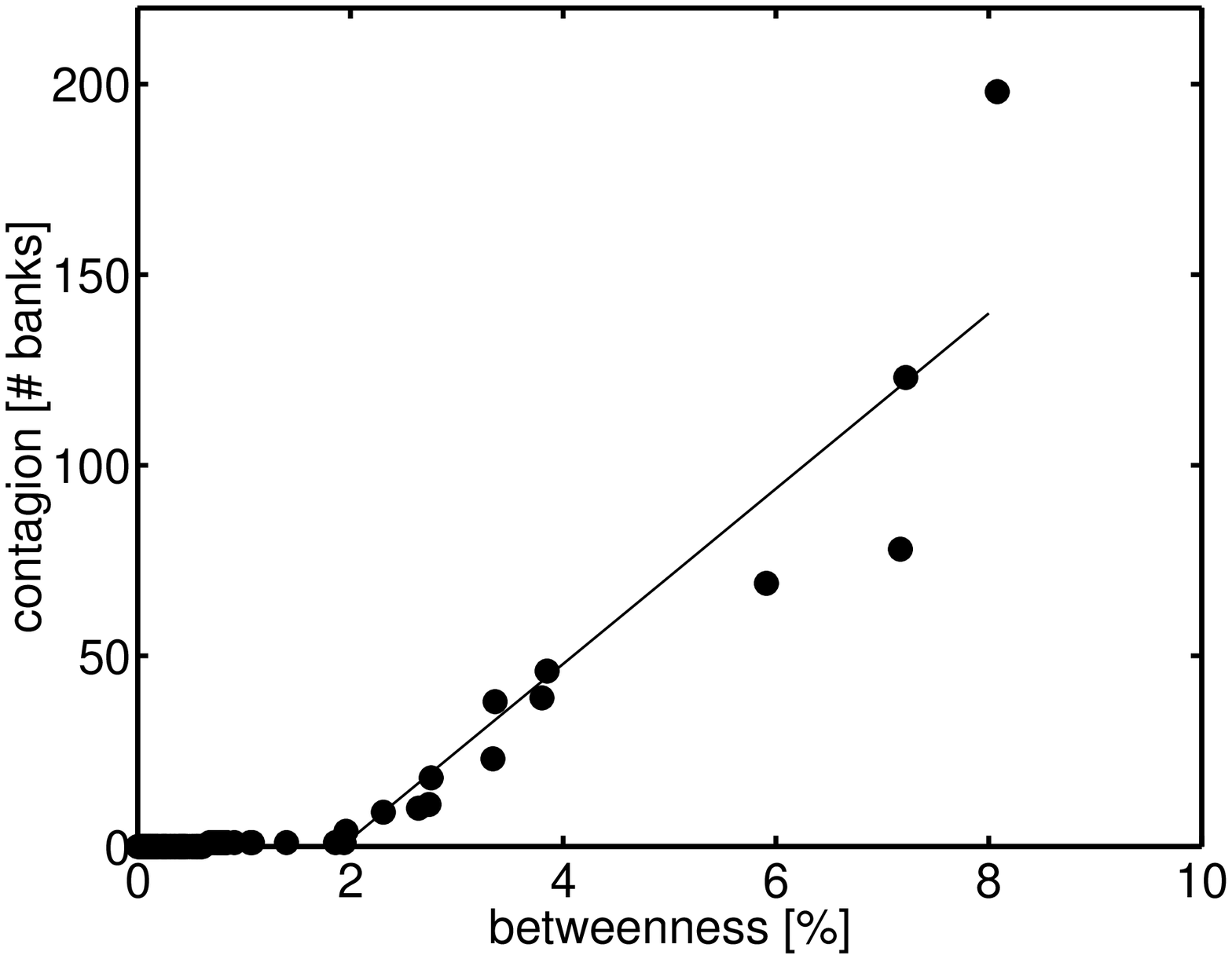} \\
\hspace{-0cm} {\large (a)} &  {\large (b)} & {\large (c)}  \\
\end{tabular}
\end{center}
\vspace{-0.6cm}
\caption{
(a) Community structure of the Austrian interbank market 
network from the same data ($A^w_{\rm clip}$). The dissimilarity 
index is a measure of the ''differentness'' of the clusters.
(b) Empirical degree distribution of the interbank 
connection network. Histograms are from 
aggregated data from the 10 datasets.
(c) Contagion impact as a function of relative node betweenness for the 
$L$ matrix. 
Below a value of $B(i)$  of 0.6 no contagion impact is found.
}
\label{pajek} 
\vspace{-0.7cm}
\end{figure}
The result for the community structure obtained from 
one representative data set is shown in Fig. \ref{pajek} a. 
The algorithm associates banks with their corresponding  sectors, like
R, VB, and S. 
For banks which in reality are not structured in a strong hierarchical 
way, such as  banks in the SP, JS, SM, HCL sectors, no significant 
community structure is expected. By the algorithm these banks are 
grouped together in a cluster called 'other'. 
The Raiffeisen sector, which has a  sub-structure in the federal 
states, is further grouped into clusters which are clearly identified 
as R banks within one of the eight federal states (B,St,K,V,T,N,O,S). 
In Fig. \ref{pajek} a these clusters are marked as e.g. 'RS', 
'R' indicating the Raiffeisen sector, and 'S' the state of 
Salzburg. 
Overall, there were 31 mis-specifications into wrong clusters within
the total $N=883$ banks, which is a mis-specification rate of 3.5 \%, 
demonstrating the quality of the dissimilarity algorithm and -- more 
importantly -- the quality of the entropy approach to reconstruct 
the matrix $L$. 

Like many real world networks, the degree distribution 
of the interbank market follows a power law for the 
tail of graph $A$, Fig. \ref{pajek} b.  The exponent is 
$\gamma_{tail}(A)=2.01$.
We have checked that the distribution for the low degrees 
is almost entirely dominated by banks of the R sector. 
Typically in the R sector most small agricultural banks have links to their
federal state head institution and very few contacts with other banks, 
leading to a strong hierarchical structure.
This hierarchical 
structure is perfectly reflected by the 
small scaling exponents. 
Betweenness is a measure of centrality that considers 
the position of nodes in-between the shortest paths (geodesics) 
that link any other nodes of the network. 
Let $g_{jk}$ be the number of geodesics linking nodes $j$ and $k$. 
If all geodesics are equally likely to be chosen 
the probability of taking 
one of them is $1/g_{jk}$. The probability that a particular node $i$ 
lies on the geodesics between $j$ and $k$ is denoted by $g_{jik}$. The 
betweenness $B$ for node $i$ is defined as the sum of these 
probabilities over all pairs of nodes not including node $i$. 
$B(i)= M \sum_{j,k} g_{jik}/ g_{jk}$, 
$M$ being some normalization constant. 
$B$ has a minimum value of zero when $i$ falls on no geodesics 
and a maximum at $(N-1) (N-2)/2$, which is the number of
pairs of nodes not including $i$. We use a relative 
version, i.e. $M$ is such  that the sum of $B(i)$ over all $N$ nodes  
adds up to 100$\%$. 
Finally, the average path length 
in the (undirected) interbank connection network $A$ is 
$\bar{\ell}(A)= 2.26\pm0.03$. 
$A$ is unweighted, meaning $A_{ij}=1$ for $A^w_{ij}\neq=0$ 
and   $A_{ij}=0$ else. 
From these results the Austrian interbank 
network looks like a very small world with about three degrees of separation.
This result looks natural in the light of the community structure
described earlier. The two and three tier organization with 
head institutions and sub-institutions apparently leads 
to short interbank distances via the upper tier
of the banking system and thus to a low degree of separation.

The framework here is a model of a banking system based on a detailed
description of the structure of interbank exposures $L$ 
\cite{elsinger02}.  The model
explains the feasible payment flows between banks endogenously from the  
given structure of interbank liabilities, net values of the banks
arising from all other bank activities and an assumption about the
resolution of insolvency for different random draws from a distribution
of risk-factor changes, such as interest rate, foreign exchange 
rate and stockmarket
changes, as well as changes in default frequencies for corporate loans. 
We expose the banks' financial
positions apart from interbank relations to interest rate, exchange
rate, stock market and business cycle shocks. For each state of the
world, the network model uniquely determines endogenously actual, feasible
interbank payment flows. Taking the feedback between banks from mutual
credit exposures and mutual exposures to aggregate shocks explicitly
into account we can calculate default frequencies of individual banks
across states. The endogenously determined vector of feasible payments
between banks also determines the recovery rates of banks with
exposures to an insolvent counterparty. We are able to distinguish
bank defaults that arise directly as a consequence of movements in the
risk factors and defaults which arise indirectly because of contagion.
The model therefore yields a decomposition into fundamental and
contagious defaults. 
Risk scenarios are
created by exposing those positions on the bank balance sheet that are not
part of the interbank business to interest rate, exchange rate, stock
market and loan loss shocks. 
\begin{figure}[t]
\begin{center}
\includegraphics[height=65mm]{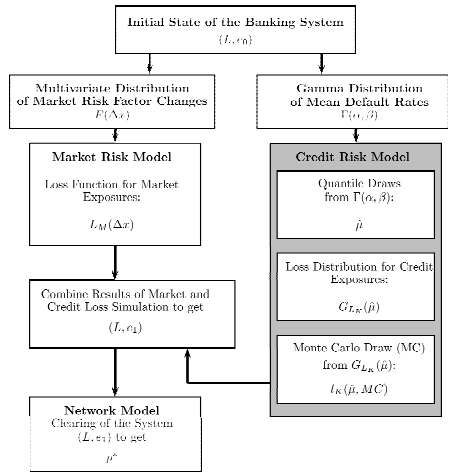}
\end{center}
\vspace{-0.7cm}
\caption{
 Schematic diagram of the  default flow model: At the initial state 
each bank is fully characterized by its wealth  $e_0$
and its liabilities to other banks $L$. We expose 
each bank to stochastic shocks which model the market risk and 
the credit risk separately. 
The introduced risk changes the 
positions of the banks to new levels $e_1$.
The crucial step is the clearing 
of the system where the structure of $L$ becomes important. The 
result of the clearing is a vector $p_i^*$ of payments bank $i$  
has to pay to the other banks in the system. 
If a component of this 
vector becomes less than the obligations $d_i$ to pay to all 
the other banks $d_i=\sum_j L_{ij}$ bank $i$ is insolvent.
}
\label{flow} 
\vspace{-1.0cm}
\end{figure}
In order to do so we undertake a historic
simulation using market data, except for the loan losses where we
employ a credit risk model. In the scenario part we use data from
Datastream, the major loans register, as well as statistics of
insolvency rates in various industry branches from the Austrian rating
agency (Kreditschutzverband von $1870$). For each scenario the estimated
matrix of bilateral exposures $L$ and the income positions determine via
the network model a unique vector of feasible interbank payments and
thus a pattern of insolvency. 
The basic idea is to determine the feasible flows of 
funds in the banking network that can occur after a 
realization of risk factor changes by a clearing
 procedure in which creditors that can not fulfill 
their payment promises ($L$) are proportionally rationed. 
One can show that a unique clearing vector always exists 
under mild regularity conditions \cite{eisenberg02}. 
The clearing vector can be found constructively 
as follows: for the given portfolio positions of banks $e_i$ and 
the given risk factor changes we assume that every bank 
would pay its interbank liabilities as specified by $L$. If 
under this assumption one bank has a negative net value 
it is insolvent and its creditors receive 
a proportional claim on the remaining value of the insolvent 
bank. All banks with positive value are assumed to honor 
their promises fully. In a second iteration it can 
happen that under the new vector of feasible payment 
flows banks that were previously solvent are now insolvent 
because the value of their interbank claims is reduced 
impairing their capacity to pay. One can show that this 
procedure converges to a unique 
clearing payment vector $p^*$ for the system as a whole. From 
the clearing vector found in this way one can directly 
read three pieces of information: first all banks that 
have a component in the vector that is smaller than the 
sum of their promises as specified by $L$ are insolvent. 
The loss given default can be determined because it 
requires only a comparison between what has been promised 
(given by $L$) and what value can actually be payed 
(given by the clearing payment vector). Furthermore,  
insolvency cases can be distinguished by their occurrence 
in the clearing algorithm. Banks that are insolvent in 
the first round are fundamentally insolvent, others are 
insolvent because the value of interbank claims has been 
reduced by the insolvency of others. Their insolvency can 
therefore be interpreted as contagious.
The analysis of these data then allows us 
to assess the risk exposure -- in particular for defaults -- 
of all banks at a system level. The details of the model are described in 
\cite{elsinger02}. 
\begin{table}[t]
\caption{
Simulation results for probabilities of 
fundamental and contagious defaults. 
A fundamental default is due to losses 
arising from exposures to market   
and non-bank credit risk. Contagious 
defaults are triggered by the default 
of another bank that cannot fulfill its promises.  
The probability that only fundamental 
defaults occur is shown as well as the probability 
that fundamental and contagious 
defaults are observed.}
\vspace{-0.1cm}
\small
\begin{center}
\begin{tabular}{|l|rrr|} \hline
Fundamental Defaults & $\,$  No Contagion & $\,$ Contagion & Total\\ \hline
0-10    &93.38\%&0.01\%&93.39\%\\
11 to 20& 2.82\%&0.40\%&3.22\%\\
21 to 30& 0.11\%&1.04\%&1.15\%\\
31 to 40& 0.00\%&0.40\%&0.40\%\\
41 to 50& 0.00\%&0.53\%&0.53\%\\
Total&96.31\%&3.69\%&100.00\%\\ \hline
\end{tabular}
\end{center}
\vspace{-0.6cm}
\label{table}
\end{table}
\section{Results and Conclusions}
The given banking system is very stable and default
events that could be classified as a "systemic crisis" are unlikely.
We find that the mean default probability of an Austrian bank to be
below one percent. Perhaps the most interesting finding is that only a
small fraction of bank defaults can be interpreted as
\emph{contagious}. The vast majority of defaults is a direct
consequence of macroeconomic shocks. More specifically, we
find the median endogenous recovery rates to be $66\%$, and we show that
the given banking system is quite stable to shocks from losses in
foreign counterparty exposure and we find no clear evidence that the
interbank market either increases correlations among banks or enables
banks to diversify risk. Using our model as a simulation tool, we show
that market share in the interbank market alone is not a good
predictor of the relevance of a bank for the banking system in terms
of contagion risk.
Table \ref{table} shows the decomposition of 
simulation scenarios with and without contagion following a 
fundamental default of a given number of banks. 
The simulation 
is run under the assumption that there is a recovery rate of 
50\% of loans to non banks and that the full value 
of an insolvent institution is transferred to the creditors. 
Clearing is done after the netting of interbank claims in $L$.


Finally, we ask the question of the impact of 
individual bank defaults on other banks. 
More specifically, if one particular 
bank becomes erased from the network, how many other 
banks become insolvent due to this event? 
We call this conditional {\it contagion impact} on 
default of a specific bank. 
This is similar in spirit to looking at 
avalanche distributions
triggered by controlled toggling of grains in a sandpile. 
We observe the portfolio positions $e_i$ of banks 
from our database. Instead of simulating risk 
factor changes that are applied to this portfolio
 and are then cleared in a second step we artificially 
eliminate the funds of each bank one at a time, 
clear the system, and count the number of contagious 
(induced) defaults.
This is repeated $N$ times so that each bank becomes 
removed, and all the others are present. 
We find that only 13 banks -- when defaulting  --
drag more than one other bank into default.   
There are 16 banks which will cause one single 
default of one other bank. A natural guess 
would be to relate the contagion impact of a 
specific bank to its role in the network. 
Amongst many possible measures, we find that the 
betweenness of the defaulting bank is directly 
related to the contagion impact. This is shown in 
Fig. \ref{pajek} c, where a linear relation 
between the betweenness $B(i)$ and the contagion impact
is suggested for $ B(i)>2$.  


In this work we combined the knowledge of the detailed structure 
of a real world banking network with an economic model, 
which allows to estimate the functional stability and 
robustness of the financial network. 
This model adds a dynamical component to the ''static'' liability 
matrix, which is nothing else but the promise for future 
financial flows between the banks. By stochastically varying 
external parameters of the system like interest rate shocks, 
we can follow  the flow of financial transactions, and in particular 
can scan for defaults occurring in the system, due to insolvency. 
The results of this work is that the system seems to be relatively 
robust, and the probability for a contagious spread  over the whole 
network is very small. However,  there are several key banks 
(not including the Central bank), which upon their default, 
will lead to a considerable number of other banks defaulting as well. 
We showed that these key banks can be reliably identified by the 
vertex betweenness.  We think that the existence of a threshold 
in the variable $B(i)$ in a ''quasi'' scale-free network, combined with  
complex but realistic contagion dynamics is a remarkable finding which 
is worthwhile to examine further.

\end{document}